\documentclass{PoS}
\def\cmm{\,{\rm cm^{-2}}}
\def\lumdens{\,{\rm ergs\,Mpc^{-3}\,s^{-1}\,Hz^{-1}}}
\def\uvunits{\,{\rm ergs\,cm^{-2}\,s^{-1}\,Hz^{-1}\,sr^{-1}}}

\title{New Population Synthesis Models of the Cosmic UV Background}

\ShortTitle{UV background}

\author{\speaker{Francesco Haardt}\\
        Dipartimento di Fisica \& Matematica, Universit\`a dell'Insubria, Como, Italy\\ 
        Department of Astronomy \& Astrophysics, University of California, Santa Cruz (CA), USA\\
        E-mail: \email{haardt@uninsubria.it}}
        
\author{Piero Madau \\
        Department of Astronomy \& Astrophysics, University of California, Santa Cruz (CA), USA\\
        E-mail: \email{pmadau@ucolick.org}}
        
\abstract{We present improved synthesis models of the evolving spectrum of the UV/X-ray diffuse 
background, updating and extending our previous results. A ``minimal cosmic reionization model" is presented 
in which the galaxy UV 
emissivity traces recent determinations of the cosmic history of star formation,
the luminosity-weighted escape fraction of hydrogen-ionizing radiation increases rapidly with redshift, from 1\% 
at $z=3$ to 50\% at $z=8.5$, the clumping factor of the intergalactic medium is $C_{\rm IGM}=3$,
and Population III stars make a negligible contribution to the metagalactic flux. The model provides a good fit to the 
hydrogen-ionization rates inferred from flux decrement measurements, 
and yields an optical depth to Thomson scattering, $\tau_e=0.085$, that is in agreement with {\it WMAP} results.}

\FullConference{25th Texas Symposium on Relativistic Astrophysics - TEXAS 2010\\
		December 06-10, 2010\\
		Heidelberg, Germany}

\begin{document}

\section{Introduction}
The reionization of the all-pervading intergalactic medium (IGM) is a landmark event in the history 
of cosmological structure formation. Studies of Gunn-Peterson absorption in the spectra of distant quasars 
show that hydrogen is highly photoionized out to redshift $z > 6$, while polarization data from the {\it Wilkinson Microwave Anisotropy Probe (WMAP)} constrain the 
redshift of a sudden reionization event to be significantly higher, $z= 11.0 \pm 1.4$. 
It is generally thought that the IGM is kept ionized by the integrated UV emission from active nuclei and 
star-forming galaxies, but the relative contributions of these sources as a function of epoch are poorly 
known. In this contribution we describe a new version of the code CUBA, aimed at solving the radiative transfer of ionizing background radiation in a clumpy, expanding medium. We will use it 
to compute improved synthesis models of the UV/X-ray cosmic background spectrum and evolution, combining, 
updating, and extending many of our previous results in this field. For details and relevant references 
see Haardt \& Madau (2011).  

\section{Cosmological radiative transfer}
The equation 
of cosmological radiative transfer describing the time evolution of the 
space- and angle-averaged monochromatic intensity $J_\nu$ is
\begin{equation}
\left({\partial \over \partial t}-\nu H {\partial \over \partial \nu}\right)J_\nu+3HJ_\nu=
- c\kappa_\nu J_\nu + {c\over 4\pi}\epsilon_\nu, \label{eq:rad}
\end{equation}
where $H(z)$ is the Hubble parameter, $c$ the speed of the light, $\kappa_\nu$ is the 
absorption coefficient, and $\epsilon_\nu$ the proper volume emissivity. The integration of 
equation (\ref{eq:rad}) gives the background intensity at the observed frequency
$\nu_o$, as seen by an observer at redshift $z_o$, 
\begin{equation}
J_{\nu_o}(z_o)={c\over 4\pi}\int_{z_o}^{\infty}\, |dt/dz| dz
{(1+z_o)^3 \over (1+z)^3} \epsilon_\nu(z) e^{-\bar\tau},
\label{Jnu}
\end{equation}
where $\nu=\nu_o(1+z)/(1+z_o)$, $|dt/dz|=H^{-1}(1+z)^{-1}$, and $\bar\tau\equiv -\ln \langle e^{-\tau}\rangle$ 
is the effective absorption optical depth of a clumpy IGM.

\subsection{IGM absorption}
The effective opacity of the IGM has traditionally been one of the main uncertainties affecting calculations
of the UV background. Our improved model uses a piecewise power-law parameterization for the distribution of 
absorbers along the line of sight, 
\begin{equation}
f(N_{\rm HI},z)=A\,N_{\rm HI}^{-\beta}(1+z)^{\gamma},
\label{eq:ladis}
\end{equation}
and is designed to reproduce accurately a number of recent observations. 

\begin{itemize}

\item Over the column density range 
$10^{11}<N_{\rm HI}<10^{15}\,\cmm$, we use $(A,\beta,\gamma)=(1.2\times 10^7,1.5,3.0)$, where the normalization 
$A$ is expressed in units of cm$^{-2(\beta-1)}$, and $\beta=1.5$.

\item At the other end of the column density distribution, with a power-law exponent $\beta=2$ down to a break column
of $N_{\rm HI}=10^{21.55}\,\cmm$, and with an incidence per unit redshift $\propto (1+z)^{1.27}$, 
the parameters for the dumped Ly$\alpha$ systems (DLAs) becomes $(A,\beta,\gamma)=(8.7\times 10^{18},2,1.27)$.

\item For absorbers with $10^{19}<N_{\rm HI}<10^{20.3}\,\cmm$ (the so-called ``super Lyman-limit systems", or SLLSs), we use 
$(A,\beta,\gamma)=(0.45,1.05,1.27)$.

\item There is obviously a significant mismatch between the power-law exponent for the Ly$\alpha$ clouds ($\gamma=3$) and the 
SLLSs ($\gamma=1.27$). Continuity then requires the shape of $f(N_{\rm HI},z)$ to change with redshift over the colum 
density range of the Lyman-limit systems (LLSs), $10^{17.5}<N_{\rm HI}<10^{19}\,\cmm$. In this interval of column densities 
we match the distribution function with a power law of redshift-dependent slope. The procedure yields the slopes 
$\beta=0.47,0.61,0.72,0.82$ at redshifts $z=2,3,4,5$, respectively.

\item The above parameterizations reproduce well the observations 
at $2< z < 5$. At low redshift, however, {\it HST} 
data show that the forest undergoes a much slower evolution. We take $\gamma=0.16$ in the interval
 $0<z<z_{\rm low}$ and $dN/dz=34.7$ at $z=0$ above 
an equivalent width of 0.24 \AA. We derive $(A,\beta,\gamma)=(1.73\times 10^8,1.5,0.16)$ for $10^{11}<N_{\rm HI}<10^{15}\,\cmm$ and
$(A,\beta,\gamma)=(5.49\times 10^{15},2,0.16)$ for $10^{15}<N_{\rm HI}<10^{17.5}\,\cmm$ at all redshifts
below $z_{\rm low}=1.56$. We use a broken power-law for the redshift distribution of the SLLSs and DLAs as well; assuming that the 
same $\gamma=0.16$ slope and transition redshift $z_{\rm low}$ inferred for the forest also hold in the case of the thicker absorbers, we 
derive a normalization at $z<z_{\rm low}$ of $A=1.28$ for the SLLSs and $A=2.47\times 10^{19}$ for the DLAs. 
This yields $dN/dz=0.74$ absorbers above $N_{\rm HI}=10^{17.2}\,\cmm$ at $\langle z\rangle =0.69$.

\item Above $z=5.5$ we assume for the forest  
the values $(A,\beta,\gamma)=(29.5,1.5,9.9)$ ($10^{11}<N_{\rm HI}<10^{15}\,\cmm$) and
$(A,\beta,\gamma)=(9.35\times 10^8,2,9.9)$ ($10^{15}<N_{\rm HI}<10^{17.5}\,\cmm$) above redshift 5.5. 

\end{itemize}

\subsection{Source emissivity}
The emissivity is due to several contributing terms:

\begin{itemize}

\item The background photons absorbed through a Lyman series resonance cause a radiative cascade that 
ultimately terminates either in a Ly$\alpha$ photon or in two-photon $2s \rightarrow 1s$ continuum decay. 
We use the detailed photoionization structure of absorbing systems to    
calculate the reprocessing of background LyC radiation by the clumpy IGM via atomic recombination processes. We 
include recombinations from the continuum to the ground state of  HI, HeI, and HeII, as well as HeII Balmer, two-photon, and 
Ly$\alpha$ emission.

\item The adopted quasar comoving emissivity at 1 Ryd, $\epsilon_{912}(z)/(1+z)^3$, is 
\begin{equation}
{\epsilon_{912}(z)\over (1+z)^3}=(10^{24.6}\,\lumdens)\,(1+z)^{4.68}\,{\exp(-0.28z)\over \exp(1.77z)+26.3},
\end{equation}
which closely fits the observational results in the range $1<z<5.7$ under the assumption of pure luminosity evolution.   
The poorly known faint-end slope of the quasar luminosity function at high redshift, incompleteness corrections, 
as well as the uncertain spectral energy distribution (SED) in the UV, all contribute to the large apparent discrepancies between 
different authors. We use the functional form given above together 
with the broken power-law SED with  $L_\nu \propto \nu^{-0.44}$ for $1300\,{\rm \AA}< \lambda <5000\,{\rm \AA}$, and 
$L_\nu \propto \nu^{-1.57}$ for $\lambda < 1300\,{\rm \AA}$.

\item Star-forming galaxies are expected to play a dominant role as sources of hydrogen-ionizing radiation at $z>3$ 
as the quasar population declines with lookback time. To compute the LyC emissivity from galaxies at all epochs, we start 
with an empirical determination of the star formation history of the universe.
We adopt the far-UV (FUV, 1500 \AA) published luminosity functions in the redshift range $0<z<9$, 
integrated down to $L_{\rm min}=0.01\,L_*$ using Schechter function fits with parameters 
$(\phi_*, L_*, \alpha)$ to compute the dust-reddened galaxy FUV luminosity density $\rho_{\rm FUV}$. Dust attenuation is treated using a 
Calzetti extinction law normalized at 1500 \AA. Finally, the dust-corrected luminosity densities are smoothed with an approximating 
function and then compared with the results of spectral population synthesis models provided by the GALEXEV library.  

\end{itemize}

\section{The Spectrum of the UV background}

Figure 1 shows the quasar-only background spectrum generated by an upgraded version of our radiative transfer code CUBA, 
using the formalism and parameters described above. CUBA solves the radiative transfer equation by iteration, as its right-hand term 
implicitly contains $J$ in the recombination emissivity and in the effective helium opacity. Physically, this simply means that the 
metagalactic UV flux depends on the ionization state of intervening absorbers, which is in turn determined by background radiation.
For comparison, we have also plotted the background spectrum from our old models. 
The new models are characterized by a lower UV flux (by as much as a 
factor of 3 at 1 Ryd and $z=3$), smaller spectral breaks from HI and HeII LyC absorption, a sawtooth modulation by the Lyman series of HI and
HeII that becomes more and more substantial with redshift. 

Figure 2 shows the full background (i.e., quasar plus galaxies) compared to the quasar-only spectrum. The much softer emissivity, linked to HI 
absorption fixed at the observed level, causes a much deeper HeII absorption trough, and much deeper HeII Ly series absorption lines. 

Figure 3 shows our predicted HI ionization rate compared to several measurements. The full model provides a good fit to the 
hydrogen-ionization rates inferred from flux decrement measurements, predicts that cosmological HII regions 
overlap at redshift 7.5, and yields an optical depth to Thomson scattering, $\tau_e=0.085$, that 
is in agreement with {\it WMAP} results.

\begin{figure}
\includegraphics[width=.8\textwidth]{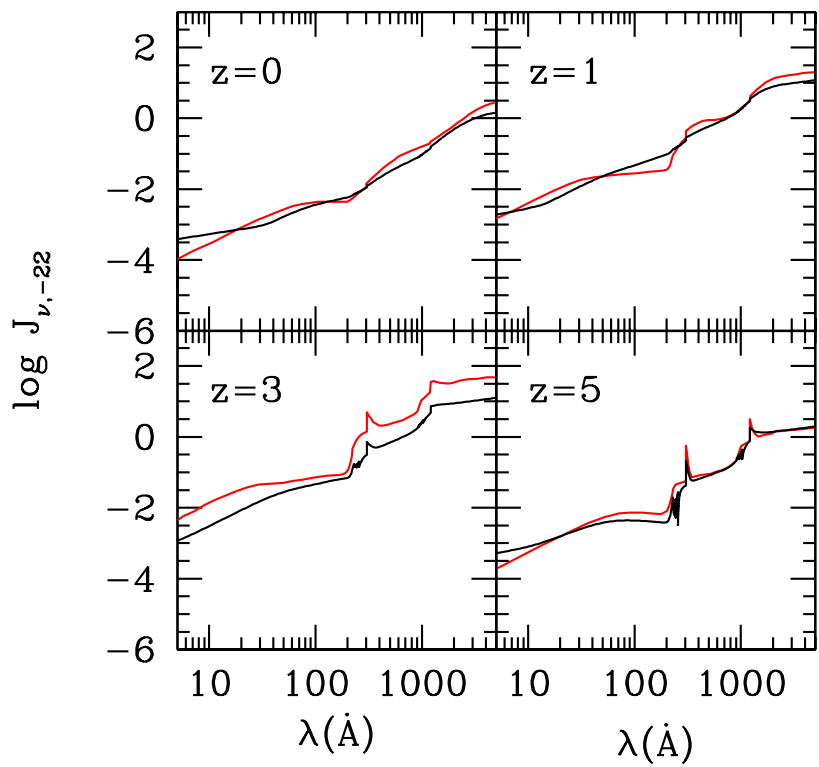}
\caption{The cosmic background spectrum from quasars only between 5 \AA\ and 5,000 \AA\ at epochs $z=0, 1, 3, $and 5. The new 
models ({\it black curves}) are compared with the old results of Haardt \& Madau (1996, {\it red curves}). The intensity $J_\nu$ is expressed in 
units of $10^{-22}\,\uvunits$.}
\end{figure}

\begin{figure}
\includegraphics[width=.8\textwidth]{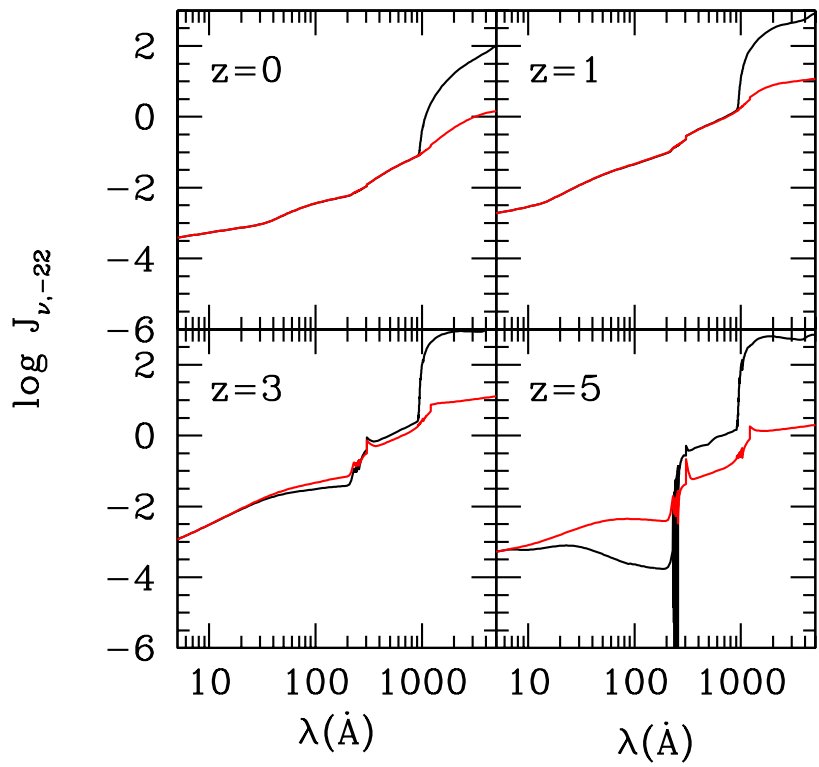}
\caption{Same as above, but now the {\it red line} is the new quasar-only background, compared to the full (quasars plus galaxies) model ({\it black 
line}).}
\end{figure}

\begin{figure}
\includegraphics[width=.8\textwidth]{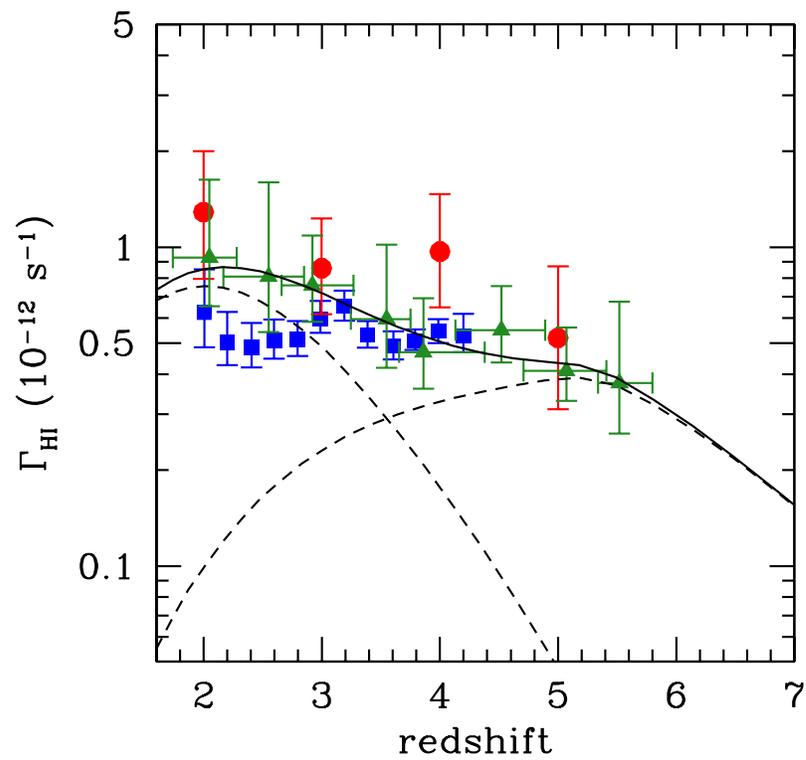}
\caption{Hydrogen photoionization rate as a function of redshift. {\it Solid curve:} our full model. {\it Dashed curves:} separate contributions from 
quasars (leftmost curve) and galaxies (rightmost curve). Different data points refer to various empirical measurements from the Ly$\alpha$ forest effective opacity (see Haardt \& Madau 2011 for full references).}
\end{figure}

\end{document}